\documentclass[slac_one]{revtex4}
\usepackage{graphicx}
\usepackage{fancyhdr}
\def \dsp{\displaystyle}
\def \sss{\scriptscriptstyle}

\def\Akstpi{\big|A_{{\sss\! K^{\!*}\!\!\pi}}\big|}

\def\Akpipi{\big|A_{{\sss\!K_{\!s}\!\pi\!\pi}}\big|} 
\def\kstpi{{\sss\! K^{\!*}\!\!\pi}}

\def\kpi{{\sss K\!\pi}}
\def\kk{{\sss K\!K}}

\def\barp{{\raise.35ex\hbox{$~{\sss (}$}}--{\raise.35ex\hbox{${\sss )}$}}}
\def\Dbarp{\hbox{$D$\kern-1.2em\raise1.4ex\hbox{\barp}}}
\def\Bar#1{\hbox{$#1$\kern-1.65em\raise1.6ex\hbox{\barp}}~}
\def\Over#1{\hbox{$#1$\kern-0.7em\raise1.6ex\hbox{--}}\kern0.15em}
\def\bBar#1{\hbox{$#1$\kern-1.2em\raise1.4ex\hbox{{\raise.35ex\hbox{$~{\sss (}$}}--{\raise.35ex\hbox{${\sss )}$}}}}}

\begin{document}

\title{Determining the Neutral D Mixing Parameters}

%

\author{Nita Sinha}
\affiliation{The Institute of Mathematical Sciences, Chennai, India}

\begin{abstract}
Evidence for mixing in the neutral $D$ meson system has recently been
reported.  Assuming negligible CP violation, non-vanishing width and mass
differences between the two neutral $D$ mass eigenstates has been found.
Theoretical predictions of these are rather difficult, obscuring detection
of New Physics contributions. However, the observation of CP violation in
the $D$ system would be a good signal of New Physics. We briefly describe the
formalism that describes the neutral $D$ decay and mixing, and present a
method to determine all the mixing parameters accurately allowing for
arbitrary CP violation.
\end{abstract}

\maketitle

\thispagestyle{fancy}

\section{INTRODUCTION} 
Neutral meson mixing is a flavor changing process, resulting in a
flavor change $\Delta F=2$ and within the Standard Model it occurs via
box diagrams with internal quarks and $W$ bosons. While the charged
current flavor violating interactions can appear at the tree level,
flavor changing neutral current (FCNC) interactions are feasible only
at the loop level. Hence these interactions have an important role to
play in the search for Physics beyond the Standard Model (SM), or New
Physics(NP), as the new particles could appear virtually in the loops.
One hopes to constrain NP by a measurement of these FCNC processes.

Untill March 2007, neutral meson mixing had only been seen in the down
type mesons: $K$(in 1956), $B_d$(in 1987) and $B_s$(in 2006). The parameter
$\epsilon_K$ in $K^0-\bar{K}^0$ mixing played a constraining role in
building of models of NP. The measured mixing in the $B_d$ and $B_s$
mesons, being consistent with SM predictions has also resulted in
constraining various NP models. $D$ meson is the only up type meson
where mixing is possible. Evidence for mixing in the neutral $D$ meson
system has recently been reported~\cite{Aubert:2007wf, Staric:2007dt, Abe:2007rd, :2007uc, Aubert:2007en} by the Belle, BaBar and CDF collaborations.
These experiments find non-vanishing width and mass differences
between the two neutral $D$ mass eigenstates assuming negligible CP
violation (CPV). The HFAG average for ICHEP08~\cite{Asner} rules out the no mixing
scenario at $9.8\sigma$. In the Standard model one expects the mixing
parameters in the D system to be small. Further, decays of the $D$
meson via tree diagrams as well as $D^0-\bar{D}^0$ mixing- due to the
negligible contribution of the internal $b$ quark in the box diagram,
essentially involve only two generations. Hence one expects that there
should be no CPV in the charm system.

\section{THEORETICAL ESTIMATES AND LOOKING FOR NEW PHYSICS}

While the internal charm quark makes the dominant contribution in the
box diagram for $K^0-\bar{K}^0$ mixing, the large top mass is responsible
for appreciable mixing in $B_d$ and $B_s$ mesons inspite of the
suppression due to the small CKM elements. In fact, mixing in the
neutral $K$ and $B_d$ systems resulted in predictions for the charm
and top quark masses respectively, before direct discovery. However,
in the $D$ meson system, the heaviest down type quark, the b quark is
not heavy enough to compensate for the large suppression due to the
small CKM factor, $|V_{ub}V_{cb}^*|^2$. Contributions from the
internal light quarks ($s$ or $d$) are dominant and hence, the mixing
parameters in the D system are expected to be small.  In the flavor
SU(3) limit one would expect mixing to exactly vanish. In fact, it has
been shown to arise only at second order in SU(3)
breaking~\cite{Falk:2001hx}.

An explicit calculation of the SM mixing parameters in the
$D$ system is very hard. This is due to the fact that the $D$ meson
mass lies in the intermediate range where neither the inclusive nor
the exclusive approaches work too well. An inclusive approach is based
on an operator product expansion in terms of matrix elements of local
operators of increasing dimensions with coefficients in powers of
$\Lambda/m_c$. However, $m_c$ is not large enough to allow the
expansion upto few terms to be accurate.  Such calculations~\cite{Bigi:2000wn} suggest
mass and width differences of order $10^{-3}$. On the other hand, in
an exclusive approach, summing over hadronic final states also fails
as $D$ is not light enough, that just a sum over few exclusive
channels could suffice. Moreover, cancellations between states within
a given SU(3) multiplet requires that the contribution of each state
be known with high precision. In Ref.~\cite{Falk:2001hx} SU(3), breaking
arising from phase space differences was studied and it was concluded
that the width difference could be at the level of one percent.

Since the theoretical predictions of the $D$ mixing parameters by
various groups vary over orders of magnitude, a clear signal of
NP from comparison of the expected and observed values of the mass and
width differences may be hard. However, an observation of CPV will
imply presence of NP~\cite{Blaylock:1995ay}, independent of hadronic uncertainties. In most
extensions of SM, the decay amplitudes and width difference are not
expected to be affected by NP, however, the mass difference could be
modified by new short distance CP violating contributions.  In our
discussion, we consider CPV only in the mixing and no direct
violation~\cite{Bergmann:2000id}.

\section{FORMALISM AND NOTATION}
Time evolution of the $D$ system is governed by the Schrodinger equation: 
\begin{equation}
i\frac{\partial}{\partial t} \left( \begin{array}{l} D^0 \\  \bar{D^0} \end{array} \right )
=  \begin{array}{c} \bf{H} \end{array}  \left( \begin{array}{r} D^0 \\  \bar{D^0} \end{array} \right )=\left( \begin{array}{cc}M_{11}-i\dsp\frac{\Gamma_{11}}{2} & M_{12}-i\dsp\frac{\Gamma_{12}}{2} \\[2ex] M_{12}^*-i\dsp\frac{\Gamma_{12
}^*}{2} & M_{22}-i\dsp\frac{\Gamma_{22}}{2} \end{array} \right )\left( \begin{array}{r} D^0 \\ \bar{D^0}  \end{array} \right )
\end{equation}
where $M_{11}=M_{22}$, $\Gamma_{11}=\Gamma_{22}$ from CPT invariance.
The neutral $D$ mass eigenstates are related to the weak eigenstates
by, $|D_{1,2}\rangle=p|D^0\rangle\pm q|\bar{D}^0\rangle~$, where
$\dsp\frac{q}{p} =~
\dsp\sqrt{\frac{M_{12}^*-i\frac{\Gamma_{12}^*}{2}}{M_{12}-i\frac{\Gamma_{12}}{2}}}~$
with $|p|^2+|q|^2=1$ is obtained by diagonalizing the Hamiltonian
matrix. The off diagonal elements of the mass matrix are due to
transitions via off-shell intermediate states while those of the decay
matrix are from on shell states and therefore constitute respectively
the the dispersive and absorptive parts of $D$ mixing. If the
magnitude of $q/p$ differs from unity and/or the weak phase
$\phi=\arg(q/p)$ is nonvanishing, this would signal $CP$ violation.

The time evolution of the states $|D^0(t)\rangle$ and
$|\bar{D}^0(t)\rangle$ which start of as pure $|D^0\rangle$ and
$|\bar{D}^0\rangle$ at $t=0$ is given by
\begin{eqnarray}
  \label{eq:Dt}
  |D^0(t)\rangle &=&
  f_{+}(t)|D^0\rangle+\frac{q}{p}f_{-}(t)|\bar{D}^0\rangle~, \\
  \label{eq:Dbar-t}
  |\bar{D}^0(t)\rangle &=&
  \frac{p}{q}f_{-}(t)|D^0\rangle+f_{+}(t)|\bar{D}^0\rangle~,
\end{eqnarray}
where,
\begin{eqnarray}
  \label{eq:f+}
  f_{+}(t)&=&e^{-iMt-\frac{\Gamma t}{2}}\cos\Big(\frac{\Delta
    M\,t}{2}-\frac{i\Delta \Gamma t}{4}\Big)~,\\
  \label{eq:f-}
  f_{-}(t)&=&-e^{-iMt-\frac{\Gamma t}{2}}\,i\,\sin\Big(\frac{\Delta
    M\,t}{2}-\frac{i\Delta \Gamma t}{4}\Big)~,
\end{eqnarray}
and $M$ and $\Gamma$ are the average mass and width of the two mass
eigenstates, also the the mass and width differences of these eigenstates are popularly
written~\cite{PDG_asner} in terms of the dimensionless variables,
\[x\equiv\dsp\frac{\Delta M}{\Gamma}=\dsp\frac{M_1-M_2}{\Gamma}
\quad\text{and}\quad y\equiv\dsp\frac{\Delta
  \Gamma}{2\,\Gamma}=\dsp\frac{\Gamma_1-\Gamma_2}{2\,\Gamma}.\]
The time evolution functions $f_\pm(t)$ are studied in the limit $x\ll
1$, $y\ll 1$ and $t\,\Gamma\ll 1$. In this limit we have
\begin{eqnarray}
  \label{eq:f-reduced}
  f_{+}(t)&=&e^{-iMt-\frac{\Gamma
      t}{2}}\Big[1-\frac{(x^2-y^2)}{8}\Gamma^2\,t^2+{\cal
    O}(\Gamma^4\,t^4)\Big]\\
  f_{-}(t)&=&\,-i\,e^{-iMt-\frac{\Gamma
      t}{2}}\Big[\frac{(x-\,i\,y)}{2}\Gamma\,t+{\cal
    O}(\Gamma^3\,t^3)\Big]
\end{eqnarray}

Hence, the time dependent decay rates for a $D^0$ decaying to a final state $f$ and
$D^0\to f$ and $\bar{D}^0\to \bar{f}$ have the form:
\begin{eqnarray}
  \label{eq:ADt2barf}
  |A(D^0(t)\to\! f)|^2 \!&= e^{-\Gamma
    t}\Big[X_{\!f}+Y_{\!f} \Gamma t+Z_{\!f}(\Gamma t)^2+\!\cdots
  \Big]~\\
  |A(\bar{D}^0(t)\to\! \bar{f})|^2\! &= e^{-\Gamma t}
  \Big[\bar{X}_{\!f}+\bar{Y}_{\!f}\Gamma
  t+\bar{Z}_{\!f}(\Gamma t)^2+\! \cdots \Big]. 
\end{eqnarray}

\section{MEASURING THE MIXING PARAMETERS}
This time dependence is utilized in determining the mixing parameters.
At $t=0$, the only term in the amplitude of decay of $D^0$ is the
direct amplitude $D^0\to f$. At any time $t>0$, there is a mixing
contribution through the sequence $D^0\to\bar{D}^0\to f$. The
interference of this mixing contribution with the direct decay,
involves the mixing parameters: $x$, $y$, $|q/p|$, $\phi$, as well as
the magnitude $r$ and strong phase $\delta$ of the ratio of the
$\bar{D}^0\to f$ and $D^0\to f$ amplitudes and plays a key role in
their measurement.

The branching ratios of the Cabibbo favored (CF) decays are large and
naively one might expect that one should use these decays to determine
the parameters. However, if $D^0\to f$ is CF, $\bar{D}^0\to f$ is
Doubly Cabibbo Suppressed (DCS) and hence the interference term is too
tiny compared to leading term in the time dependent decay
amplitude. For example, the decay rate for $D^0\to K^-\pi^+$ is,
\begin{equation}
\label{eq:CF}
\Gamma(D^0\to K^-\pi^+)= |A_{K\pi}|^2 e^{-\Gamma
    t}\Big[1+|\frac{q}{p}|r_{K\pi}\big[x \sin(\delta-\phi)+y\cos(\delta-\phi)\big]\Gamma t
+...
  \Big], 
\end{equation}
where, $A_{K\pi}\equiv A(D^0\to K^-\pi^+ )$ and the ratio of the DCS to CF amplitude is defined as: 
\begin{equation}
-r_{k\pi} e^{-i\delta_{K\pi}}\equiv\frac{A(\bar{D}^0\to K^-\pi^+)}{A(D^0\to
  K^-\pi^+ )}=\frac{A(D^0\to K^+\pi^-)}{A(D^0\to K^-\pi^+)}~. 
\end{equation}
Since $r_{k\pi}$, $x$ and $y$ are much less than unity, even the
linear term in $\Gamma\!t$ is negligible compared to the constant
term and the CF decay rate can can only be used to determine
$|A_{K\pi}|$.  In a DCS mode on the other hand the constant, linear
and quadratic terms in $\Gamma\!t$ in the time dependent decay rate
are all of the same order, allowing all three terms to be measurable.
The decay rate for the DCS mode $D^0\to K^+\pi^-$ is,
\begin{equation}
  \label{eq:DCS}
  \Gamma(D^0\to K^+\pi^-)= |A_{K\pi}|^2 r_{K\pi}^2 e^{-\Gamma
    t}\Big[1+|\frac{q}{p}|\frac{y'_{K\pi}\cos\phi-x'_{K\pi}\sin\phi}{ r_{K\pi}}\Gamma t
+|\frac{q}{p}|^2\frac{x'^2+y'^2}{4 r_{K\pi}^2}(\Gamma t)^2
  \Big]~,
\end{equation}
where, due to the presence of a relative strong phase between
the DCS and CF amplitudes, the combinations:
$x^\prime_{\kpi}=(x\cos\delta_{\kpi}+y\sin\delta_\kpi)$ and
$y^\prime_{\kpi}=(y\cos\delta_\kpi-x\sin\delta_\kpi)$ 
appear. In the linear and quadratic terms, the suppression from the
small mixing parameters is compensated by the larger $\bar{D}^0\to f$
rate.  The time dependent decay rate of $D^0\to K^+\pi^-$ as well as
its conjugate mode was used by BaBar and CDF collaborations to
determine~\cite{Aubert:2007wf, :2007uc} $y^\prime$ (from the linear term) and $x^{\prime 2}$ (from
the quadratic term), assuming CP conservation.

In the case of singly Cabibbo
suppressed (SCS) $CP$ eigenstates modes, the strong phase is
identically zero; and hence, the time dependent decay rate for these
modes, like $D^0\to K^-K^+$ reduces to the simple form:
\begin{eqnarray*}
\Gamma(D^0\to K^+K^-)= |A_{\kk}|^2  e^{-\Gamma
    t}\Big[1-|\frac{q}{p}|(y_{\kk}\cos\phi-x_{\kk}\sin\phi)\Gamma t \Big]~.
\end{eqnarray*}
Unlike the DCS modes where the term quadratic in $\Gamma\!t$ is
enhanced by the ratio of CF to DCS rates, in the SCS modes all time
dependent terms are of the same order in $\sin\theta_c$, hence
quadratic and higher terms in $\Gamma\!t$ cannot be extracted.
Assuming $|q/p|\approx 1$ and $\phi=0$, the linear term in $\Gamma\!t$
can directly measure $y$, as has been done in Ref.~\cite{Staric:2007dt}($y_{CP}$ of Ref.~\cite{Staric:2007dt} is $y$ for no CPV).  However, the time
dependent study of only the SCS CP eigenstates does not allow $x$ to
be determined, even in the limit $|q/p|\approx 1$ and $\phi=0$.

A Dalitz plot analysis~\cite{Abe:2007rd} of $D^0\to K_s\pi^+\pi^-$ has also been
performed by the Belle collaboration to determine all the mixing
parameters. But this has systematic errors associated with the
parameterization of the resonant content of the Dalitz plot and hence
is model-dependent. Measurements of $y_{CP}$ using $D^0\to K_s K^+K^-$
as well as that of $y^\prime$ and $x^{\prime 2}$ using $D^0\to K^+\pi^-\pi^0$
were also reported~\cite{Meadows} at this conference.

Since the SM estimates of $x$ and $y$ are uncertain, the current
experimental results for these parameters cannot have any clear
implications for NP. However, these results already constrain the
parameter space of various models. In supersymmetric models with
quark-squark alignment, constraints on the up-type squark matrices
have been discussed in Ref.~\cite{Nir:2007ac}. They seem
to imply squark and gluino masses above 2 TeV. A detailed analysis of
various NP models has been carried out in Ref.~\cite{Golowich:2007ka}.  Out of the
21 models considered, only 4 are ineffective in producing charm mixing
at the observed level. For the rest of the 17 models constraints on
masses and mixing parameters are obtained.

As pointed out earlier, within the SM, CPV in the $D$ system is
negligible and an observation of CPV would be a clear signal of New
Physics. Babar and Belle have looked for CPV by calculating $y^\prime$
and $x^{\prime 2}$ for $D^0$ and $\bar{D}^0$ separately. They have
also searched for CPV by measuring the difference of the decay rates
of $\bar{D}^0$ and $D^0$ to SCS CP eigenstates. No evidence of CPV has
been obtained. It is hence important to have a technique to accurately
measure the CP violating phase.

\section{DETERMINATION OF THE CP VIOLATING PHASE ALONG WITH OTHER
  MIXING PARAMETERS}

A technique to accurately determine all the mixing parameters
including the CP violating phase has been given in Ref.~\cite{Sinha:2007zz}.
It was shown that using the DCS mode $D\to
K^{*0}\pi^0$ and its conjugate modes, one can solve for all the
$D-\bar{D}$ mixing parameters.  This is possible if the
$K^{*0}/\bar{K}^{*0}$ is reconstructed both in the self tagging
$K^\pm\pi^\mp$ mode and in the CP eigenstate $K_{\sss\! S}\pi^0$ mode.

With the $K^{*0}/\bar{K}^{*0}$ reconstructed in the self tagging
$K^\pm\pi^\mp$ modes, the time dependent decay rate has a form similar
to that in Eq.~(\ref{eq:DCS}). The amplitude $\Akstpi\equiv |A(D^0\to
\bar{K}^{*0}\pi^0)|$ can easily be measured using the time integrated
rate for the CF mode $D^0\to \bar{K}^{*0}\pi^0$.  The magnitude of the
ratio of the DCS to CF amplitude can hence be determined using
constant term in time dependent decay rate. The quadratic terms in
$\Gamma\!t$ in the time dependent decay rates of $D^0\to K^{*0}\pi^0$
and its conjugate mode $\bar{D}^0\to \bar{K}^{*0}\pi^0$ readily
determine $|q/p|$ and $x^{\prime 2}+y^{\prime 2}$. The interference
terms, now involve the 4 unknown parameters: $x$, $y$, $\phi$ and
$\delta$, with a known value of $x^{\prime 2}+y^{\prime 2}$. To
determine them all, an additional observable is required.

The $K^{*0}/\bar{K}^{*0}$ in the final state could also have been
reconstructed in the neutral $K_{\sss\! S}\pi^0$ mode, this provides
the required additional observable. A unique feature of the final
state $K_{\sss\!S}\pi^0\pi^0$ is that it includes contributions from
both $K^{*0}\pi^0$ as well as $\bar{K}^{*0}\pi^0$ states; the
amplitude for this final state is thus a sum of the CF and DCS
amplitudes,
\begin{eqnarray}
  \label{eq:kspipi}
\Akpipi^2 &\equiv& |A(D^0\to K_{\sss\! S}\pi^0\pi^0)|^2 =|A_\kstpi|^2(1+r_\kstpi^2-2\,r_\kstpi\cos\delta_\kstpi)~. 
\end{eqnarray}
Since the decay mode involves two neutral pions it will not be easy to
perform a time dependent study. Hence, we consider only the time
integrated decay rate for this mode.  The amplitudes $A(D^0\to
K_{\sss\! S}\pi^0\pi^0)$ and $A(\bar{D}^0\to K_{\sss\! S}\pi^0\pi^0)$
are equal since $K_{\sss\! S}\pi^0\pi^0$ is a CP eigenstate. Hence,
the time integrated decay rate for $D^0\to K_{\sss\! S}\pi^0\pi^0$ has
the form,
\begin{eqnarray}
\int_0^\infty |A(D^0(t)\to K_{\sss\! S}\pi^0\pi^0)|^2 dt\approx |A_\kstpi|^2\big[1+\!\frac{q}{p}\!(y\,\cos\phi-x\,\sin\phi)\!
-\!2 r_{\kstpi}\cos\delta_\kstpi\big].
\end{eqnarray}
Using this along with the linear terms of the time dependent decay rates of
the self tagging modes allows a solution for $\tan^2\phi$ and for
$x/y$ with a four-fold ambiguity. $x$ and $y$ can thus be individually
determined since $x^{\prime 2}+y^{\prime 2}$ is known. The solution obtained is
finite even if $\phi=0$, with a correction term of order
$(x\cos\delta_\kstpi+y\sin\delta_\kstpi)\sin\phi$. Hence an accurate
estimation is possible, even if $\phi$ is tiny. It should be
possible to determine $|x|$, $|y|$ to order $7\times 10^{-4}$,
$4\times 10^{-4}$ respectively and $\phi$ to about $1^o$ 
at a Super-B factory with an integrated luminosity of
$50\,ab^{-1}$~.

While the SCS CP eigenstates like $D\to K^+K^-$ cannot alone be used
to determine all the mixing parameters, minimal additional
information from DCS modes makes this possible. This approach may provide the optimal method to
determine all the parameters with current data. For the SCS-CP
eigenstates, the strong phase is identically zero and the ratio $r=1$.
The coefficients of the linear term in $\Gamma\!t$, in the time
dependent decay rate is a function of 3 parameters: $x$, $y$ and
$\phi$. As pointed out earlier the quadratic and higher terms in
$\Gamma\!t$ cannot be extracted. However, if we
also include in this analysis the quadratic terms in $\Gamma\!t$
from the time dependent decay rates of DCS modes such as $K\pi$, all
the mixing parameters can be solved without approximation. Since
$x^{\prime 2}+y^{\prime 2} = x^2+y^2$, these quadratic terms readily
determine $|q/p|$ and $f^2=x^2+y^2$. Alternatively, $|q/p|$ and $f^2$
could be measured using time integrated wrong sign relative to right
sign SL decay rates. Having obtained $|q/p|$ and $f^2$, $\phi$ and
$x/y$ can easily be determined from $D\to K^+K^-$ and the solutions
have been shown to be finite even for small $\phi$.

It may be further pointed out that if information from $K^+K^-$ modes
is added to that from the $K^*\pi$ modes it helps in reducing the
ambiguities in $x$ and $y$ from four-fold to two-fold.
It has recently been proposed to use the singly Cabibbo suppressed
(SCS) $D\to K^*K$ modes to determine the mixing
parameters~\cite{Xing:2007sd,Grossman:2006jg}.  However, if $\phi$ is
zero, these methods would be feasible only if the strong phase
involved is measured elsewhere.

\section{CONCLUSIONS}
Mixing in the neutral $D$ system has been clearly established. The
Standard Model predictions involve large uncertainties, obscuring
detection of New Physics contributions. An
observation of CP violation in $D$ mixing would clearly imply New
Physics. The $D\to K^{*0}\pi^0$ modes are an example where it is
possible to measure the CP violating parameters as well as the mass
and width differences of the two $D$ meson mass-eigenstates using only
related final states, thereby reducing systematic errors.

\begin{acknowledgments}
The author would like to thank S.~Pakvasa, N.~G.~Deshpande, T.~Browder
and R.~Sinha for collaborating on work on which part of this talk is based.
\end{acknowledgments}

\end{document}